\documentclass[a4paper,12pt]{article}
\usepackage{epsfig}
\def\be#1{\begin{equation}\label{#1}}
\def\ee{\end{equation}}
\def\eq#1{(\ref{#1})}

\newcommand {\Oe}       [1]     {$$}

\newcommand {\ba}       [2]     {\be{#1} \begin{array}{#2}}
\newcommand {\ea}               {\end{array} \ee}
\newcommand {\Oa}       [2]     {$$ \begin{array}{#2}}
\def\oa{\end{array} $$}

\let\|=\over

\newcommand {\qq}   {\,,\qquad}

\newcommand {\defeq}{\stackrel{\mbox{\scriptsize def}}{=}}
\newcommand {\hence}{\quad\Rightarrow\quad}

\let\eps=\varepsilon
\let\w=\omega
\let\al=\alpha

\def\({\left(}
\def\){\right)}

\let\ap=\approx

\def\fig#1{Fig.\,#1}

\def\f{\varphi}
\def\W{\Omega}

\let\pa=\partial

\begin{document}

\baselineskip 18pt

\epsfxsize=0.45\hsize

\let\ns = \normalsize

\title{
\bf Dissociation of Diatomic Molecule by Energy-Feedback Control }
\author{
Alexander Fradkov, Anton Krivtsov, Alexander Efimov
\footnote{The authors acknowledge support of the Russian Foundation of Basic
Research (grant RFBR 99-01-00672) and of the Program of basic
research N 19 (project 1.4) of the Presidium of RAS.}
\\ \ns Institute for Problems of Mechanical Engineering,
\\ \ns Russian Academy of Sciences,
\\ \ns 61 Bolshoy ave., V.O., St.~Petersburg, 199178, Russia.
\\ \ns Fax: +7(812)321-4771, Tel: +7(812)321-4766,
\\ \ns E-mail:
alf@ipme.ru }
\date{December 26, 2001}
\maketitle

\begin{abstract}
New method for dissociation of diatomic molecule based on
nonperiodic excitation generated by energy-feedback control
mechanism is proposed. The energy-feedback control uses
frequency-energy (FE) relation of the natural oscillations to
fulfill the resonance conditions at any time of excitation.
 Efficiency of the proposed method is demonstrated by the problem
 of dissociation of hydrogen fluoride (HF) molecule. It is
shown that new method is more efficient then methods based on
constant frequency and linear chirping excitation.
\end{abstract}

\noindent {\bf Keywords:} diatomic molecule, dissociation,
control, feedback.

\section{Introduction}

During last decade a growing interest has been observed in the
control problems for molecular systems in classical and quantum
formulation \cite{CHEMPHYS01, Rabitz00, Yu97}. One of the simplest
problems of that type is dissociation problem for diatomic
molecule \cite{Yu97, Goggin88, Goggin88a, GB91, Krempl92, Liu95}.
It often serves as a benchmark for comparison of classical and
quantum calculations. In the paper \cite{Goggin88} possibilities
of dissociation a molecule by monochromatic (single frequency)
laser field have been explored for the case of hydrogen fluoride
(HF) molecule using Chirikov's resonance overlap criterion. In
\cite{Goggin88a, GB91} the case of two-frequency (two-laser)
control field was investigated. It was shown that intensity of
two-frequency field required for dissociation can be reduced
compared to one-frequency case. In \cite{Liu95} the further
reduction of the control field intensity has been demonstrated by
means of chirping (frequency modulation) the laser frequency with
constant chirping rate.

New possibilities for changing of physical and chemical properties
are provided by using feedback. In \cite{Yu97} methods based on
geometric control theory (inverse control) were proposed for
molecular systems.
%However, physical reasonability of the methods of \cite{Yu97}  is
%sometimes problematic (?).
%However, physical realization of the methods of \cite{Yu97} could
%be problematic.
In the papers \cite{F99, F99ecc} a general method
for investigation of physical systems by means of feedback
controlling actions was proposed. It was shown in \cite{F99} by
example of the problem of escape from a potential well that
feedback allows to reduce control intensity required for
overcoming a potential barrier by several orders of magnitude. It
appears that using speed-gradient method \cite{FP98, FMN99} for
design of feedback control action allows to achieve desired level
of energy by excitation of intensity inversely proportional to the
system dissipation degree. Particularly, for conservative systems
it allows to reach any energy level by (ideally) arbitrarily small
control action.

In this paper the algorithm of the feedback frequency
chirping of the controlling field frequency is proposed based on
dependence of the natural frequency of the nonlinear molecule
oscillations on its energy, further referred to as {\it frequency-energy
characteristics} or {\it frequency-energy curve (FE-curve)}.
%``skeleton curve'' of the amplitude-frequency
%characteristic).
On-line implementation of the algorithm requires measurement of
molecule energy only. This can be easily generalized for an
ensemble of molecules, replacing exact energy of the separate
molecule in the control criterion by the average energy per single
molecule. However, for control algorithm design the knowledge of
the frequency-energy relation is required. In this paper a
classical Morse potential is considered, which allows obtaining
this relation in a simple analytical form. For other potentials,
where the analytical solution is too complicated or impossible,
the frequency-energy relation can be obtained numerically.
Efficiency of the proposed algorithm  for the case of HF molecule
is illustrated by computer simulation. It is shown that the time
of dissociation can be reduced significantly when using proper
 feedback excitation.

\iffalse
In Section 2 the algorithm of the feedback frequency chirping of
the controlling field frequency is proposed based on dependence of
the natural frequency of the molecule oscillations on their
amplitude (``frequency-energy curve''). On-line implementation of the
algorithm requires measurement of  molecule energy only. However,
for control algorithm design knowledge of the ``frequency-energy curve''
depending on shape of the potential  is necessary. Efficiency of
the proposed algorithm  for the case of HF molecule is confirmed
by computer simulation. It will be shown that the dissociation time can
be significantle reduced by using feedback control algorithm

In Section 3 simulation results are presented for second algorithm
based on speed-gradient-energy approach. Speed-gradient algorithm
does not depend on the shape of the potential curve and does not
require precalculation of the ``skeleton curve''. However, it
requires measurement of the system momentum. Although for
unimolecular systems such a possibility is problematic, for
multi-molecular ensembles measuring of average momentum may be
sufficient that looks more realistic.
\fi

\section{Basic equations}

Let us consider a diatomic molecular system under the action of the
external laser field. Dynamics of such a system can be described by the
following controlled Hamiltonian~\cite{GB91,Yu97}
\be1
    H = \frac{p^2}{2m} + \Pi(r) - \mu(r) ~u(t) \,,
\ee
where $m$ is reduced mass, $p$ is momentum, $\Pi(r)$ is potential of
interatomic interaction, $\mu(r)$ is dipole moment of the molecule,
$u(t)$ is intensity of external field. The value $u(t)$ serves as control variable.
E.g. for monochromatic control field $u(t)= E \cos(\w t)$, where
$E$ and $\w$ are strength and frequency of the external
laser field. Substitution of \eq1 into Hamilton equations
%of motion
\be2
    \frac{\pa r}{\pa t} =  \frac{\pa H}{\pa p} \qq
    \frac{\pa p}{\pa t} = -\frac{\pa H}{\pa r}
\ee
yields the following equation of molecular motion
%Corresponding equation of motion is
\be{2u}
   m\ddot r = -\Pi'(r) +  \mu'(r) u(t) \,.
\ee
For the case of  harmonic excitation equations of motion is as follows:
\be3
   m\ddot r = -\Pi'(r) + E \mu'(r) \cos(\w t) \,.
\ee
Let us use Morse form of the interatomic potential
\be4
   \Pi(r) = D \(1 - e^{-\al(r-a)}  \)^2 - D
   = D \(e^{-2\al(r-a)} - 2 e^{-\al(r-a)}  \)
    \,,
\ee
where $D$ is the bond energy, $a$ is the equilibrium interatomic
distance. The corresponding force is
\be5
   f(r) = -\Pi'(r) = 2\al D \(e^{-2\al(r-a)} - e^{-\al(r-a)}  \)
    \,.
\ee
The dipole moment can be represented in the form \cite{GB91, SN79}
\be6
    \mu(r) = A r e^{-\xi r^4}
    ~\mu'(r) = A \(1 - 4\xi r^4 \) e^{-\xi r^4} \,,
\ee
where $A$ and $\xi$ are the known constants. Thus the equation of
motion \ref{3}  reads
\be{3a}
   m\ddot r = 2\al D \(e^{-2\al(r-a)} - e^{-\al(r-a)}  \)
            + E A \(1 - 4\xi r^4 \) e^{-\xi r^4} u(t) \,.
\ee
For sufficiently small $\xi$ the simplified expression for dipol potential
can be used:
\be{6a}
    \mu(r) = A r,   ~\mu'(r) = A  \,,
\ee
In the vicinity of equilibrium $r\ap a$ equation \eq{3a} with
harmonic excitation can be
reduced to the equation of linear forced oscillations
\be7
   m\ddot r = C r + E \mu'(a) \cos(\w t)
\qq
    C \defeq \Pi''(a) = 2\al^2 D
    \,,
\ee
where $C$ is linear stiffness of the bond. The natural frequency
of the linear system is
\be8
    \W_0 \defeq \sqrt{C/m} = \al\sqrt{2D/m}
\,.
\ee

\section{Control of the dissociation process}

In a nonlinear system the resonance frequency is function of the
amplitude of oscillations, or, in other words, it depends on the
energy of the oscillator. Thus the faster dissociation can be
achieved if the excitation frequency is changing (decreasing in our case)
 while the energy increases. The simplest control algorithm of
this kind can be done by linear chirping:
\be9
    \w = \W_0 - \eps t
\ee
where $\eps$ is a constant (chirping rate), which characterizes the speed
 of the frequency decrease. In this case equation of motion \eq3 takes the
form
\be{10}
   m\ddot r = -\Pi'(r) + E \mu'(r) \cos\f \qq \dot\f = \w(t) = \W_0 - \eps t \,.
\ee
(Note that in this case the representation $\f = \w t$ is
not  valid since the frequency $\w$ is not constant.

%More smart way is to change $\w$ according to the oscillator's
%energy.

%More effective dissociation can be achieved by taking advantages
%of the feedback control. Unfortunately many very effective
%feedback methods (e.g. velocity gradient) hardly can be used in
%molecular systems since they require continuous observation of the
%velocity of each atom. More realistic way is to measure energy of
%the atom oscillations. Therefore we suggest feedback control based
%on knowing relation between

More effective dissociation can be achieved by taking advantage of
the feedback control. In this case the frequency of excitation can
be changed according to the oscillator energy so that at any
amplitude of oscillations the excitation will act at the resonance
frequency. %To implement this knowledge of the relation
Let  $\W(W)$ be the natural frequency of the diatomic molecular
oscillator at the specified energy $W$. We call the function $\W(W)$
{\it frequency-energy curve} of {\it FE-curve}. It
contains important information about dynamics of the molecular motion.
The key  idea of our approach is that if the
curve $\W(W)$ is known and the energy $W = W(t)$
is observable then control algorithm
\be{11}
   m\ddot r = -\Pi'(r) + E \mu'(r) \cos\f \qq \dot\f = \W(W)
\ee
will give the desired excitation at the resonance frequency at any
 time instant.

The relation $\W(W)$ in principle can be calculated for any known
potential $\Pi(r)$ using the integral of energy, which is realized
when the external force is absent
%Indeed, for the free oscillations the integral of
%energy is realized
\be{12}
    \frac12\,m\dot r^2 + \Pi(r) = W \hence
    T = \sqrt{2m} \int\limits_{r_1}^{r_2} \frac{d r}{ \sqrt{W-\Pi(r)} } \,,
\ee
where $W$ is a constant value of energy, $T$ is period of the
oscillations, $r_1$ and $r_2$ are the minimum and the maximum
possible values of radius $r$ for the specified potential energy (the
solutions of the equation $\Pi(r)=W$). If period $T$ is known then
frequency can be obtained as $\W=2\pi/T$. Obviously integral
\eq{12} can not be calculated analytically for arbitrary
potential and  the desired relation
$\W(W)$  should be evaluated by numeric integration.

Luckily for the Morse potential
\eq4 the integral can be evaluated analytically in the following way.
%which is frequently used to describe simple molecular
%systems the integral can be calculated analytically. The Morse
%potential has the from
Substitution $\zeta = \zeta(r) = e^{\al(r-a)}$ transforms the
integral \eq{12} in the case of Morse potential to the simple form
which can be easily integrated
\be{13}
    T = \frac1\al\,\sqrt{\frac{2m}{D}}
%    \int\limits_{\zeta_1}^{\zeta_2} \frac{d\zeta}{ \sqrt{h\zeta^2+2\zeta-1} }
    \int\limits_{\zeta_1}^{\zeta_2} \frac{d\zeta}{ \sqrt{\frac{W}{D}\,\zeta^2+2\zeta-1} }
    = \frac\pi\al\,\sqrt{-\frac{2m}{W}}\,.
\ee
The above formula is valid if $W<0$. Otherwise,
the motion of the system is not periodic, since the molecule
dissociates. Equation \eq{13} provides the following simple relation between
energy and natural frequency of Morse oscillator
\be{14}
    -\frac W D = \(\frac{T_0}{T}\)^2 = \(\frac{\W}{\W_0}\)^2 \,,
\ee
%where $\W_0 = $
where $\W_0$ is defined by \eq{8}, $T_0 = 2\pi/\W_0$. The control
law \eq{11} therefore takes the form
\be{15}
   \dot\f = \W(W) = \W_0\sqrt{-\frac {W(t)} {D}} \,.
\ee
where $W(t)$ is the current value of the molecule energy.

\section{Calculations}
Efficiency of the proposed control algorithm was confirmed by
computer simulations for dissociation of the hydrogen fluoride (HF)
molecule  that has become a benchmark problem for controlled
dissociation ~\cite{Yu97, GB91, SN79}.
The parameters of the model \ref{3a},
corresponding to HF molecule~\cite{Yu97, GB91, SN79} are as
follows: $m = 1732$, $D = 0.2101$,
$\al = 1.22$, $a = 1.75$, $A = 0.4541$, $\xi = 0.0064$, $E = 0.1$.
All quantities are given in atomic units. The initial condition
for excitation is molecule equilibrium at the bottom of the
potential well: $r = a$, $\dot r = 0$. The equation was integrated
numerically by central differences method with time step $d t = T_0 / 160$
(integration with smaller time steps yields practically the same results.

\begin{figure}[htb]
a)\hspace{-1mm}
\hbox{\epsfbox{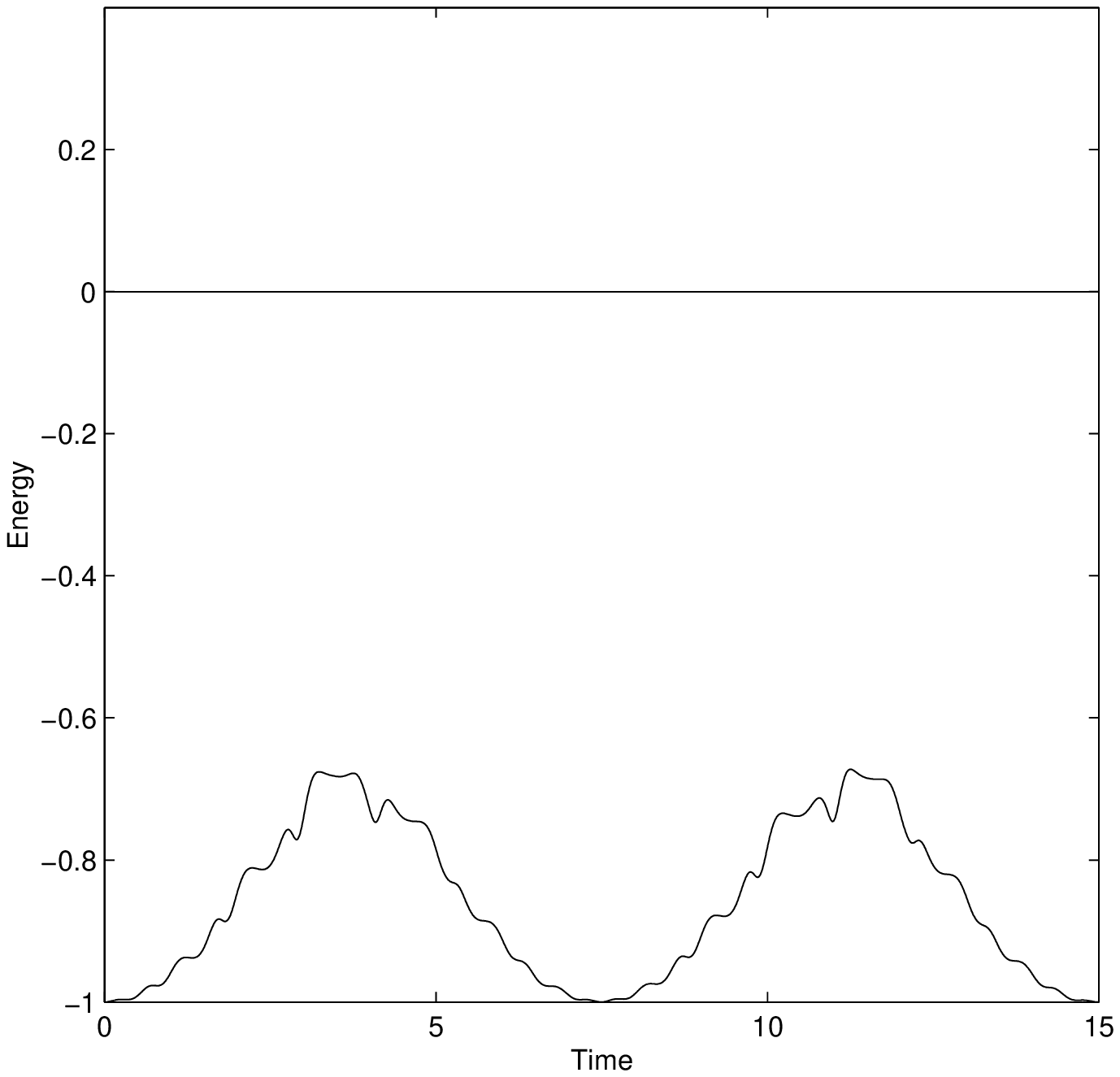}}
%\fbox{\epsfbox{fig01.eps}}
\hspace{7mm}
b)\hspace{-1mm}
\hbox{\epsfbox{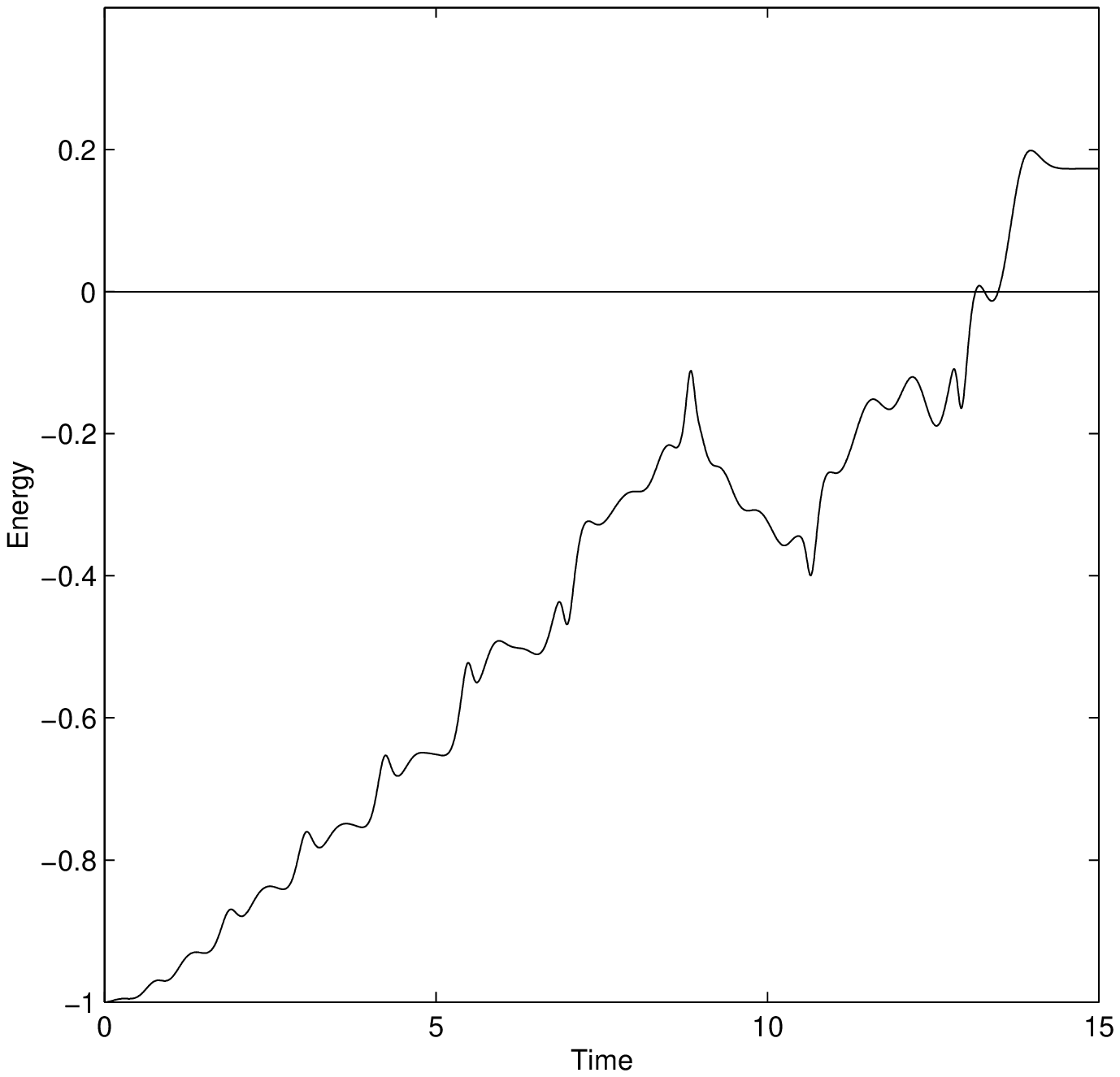}}
%\fbox{\epsfbox{fig02.eps}}
\caption{\label{fig-1}
Excitation at constant frequency: a) resonance frequency $\w =
\W_0$, b) optimal frequency $\w = 0.841\W_0$.
Energy is measured in the units of the bond energy $D$, time is in
the units of $T_0$ (natural period of linear oscillations).}
\end{figure}

The result of excitation at the resonance frequency is shown in
\fig{1a}. The graph shows that the motion is quasiperiodic and
maximum value of energy, which can be achieved by this method, is
$-0.68D$. Since the natural frequency of the molecule is
decreasing when the energy is growing, more efficient excitation can
be achieved at lower frequencies. It is even possible to
dissociate the molecule if the excitation frequency is close to
$0.841\W_0$ --- see \fig{2a}. The figure shows that at $t\ap 13.5
T_0$ the energy crosses the zero level, and at $t= 14.5 T_0$ the
energy becomes a positive constant. This practically means that
the potential energy is negligible and atoms are moving in the
opposite directions with constant velocities (total dissociation).
However this result is not structurally stable:
small variation in the excitation frequency can noticeably
increase the dissociation time and even break the dissociation.

\begin{figure}[htb]
a)\hspace{-1mm}
 \hbox{\epsfbox{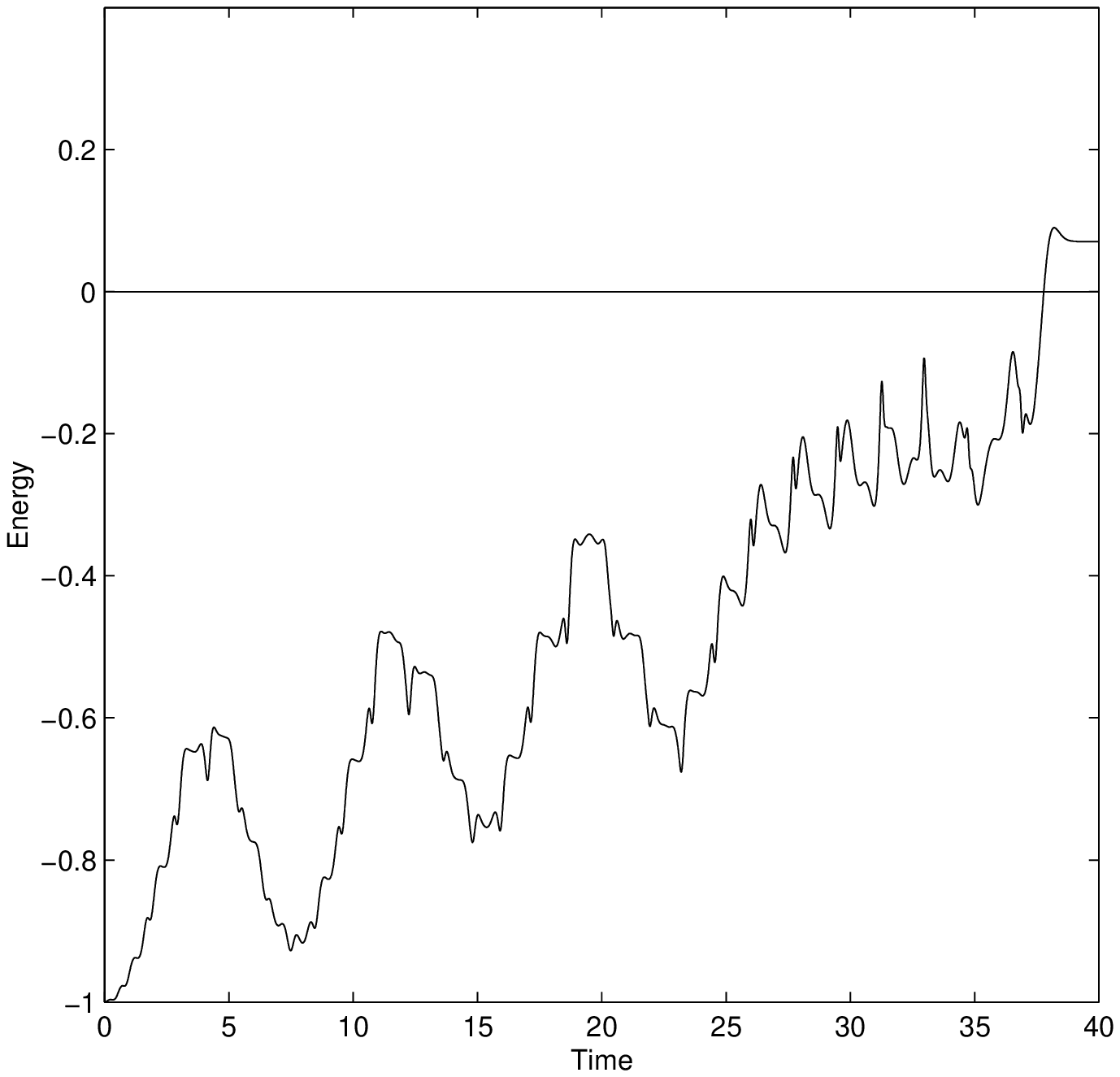}}
\hspace{7mm}
b)\hspace{-1mm}
% \hbox{\epsfbox{fig03b.eps}}
 \hbox{\epsfbox{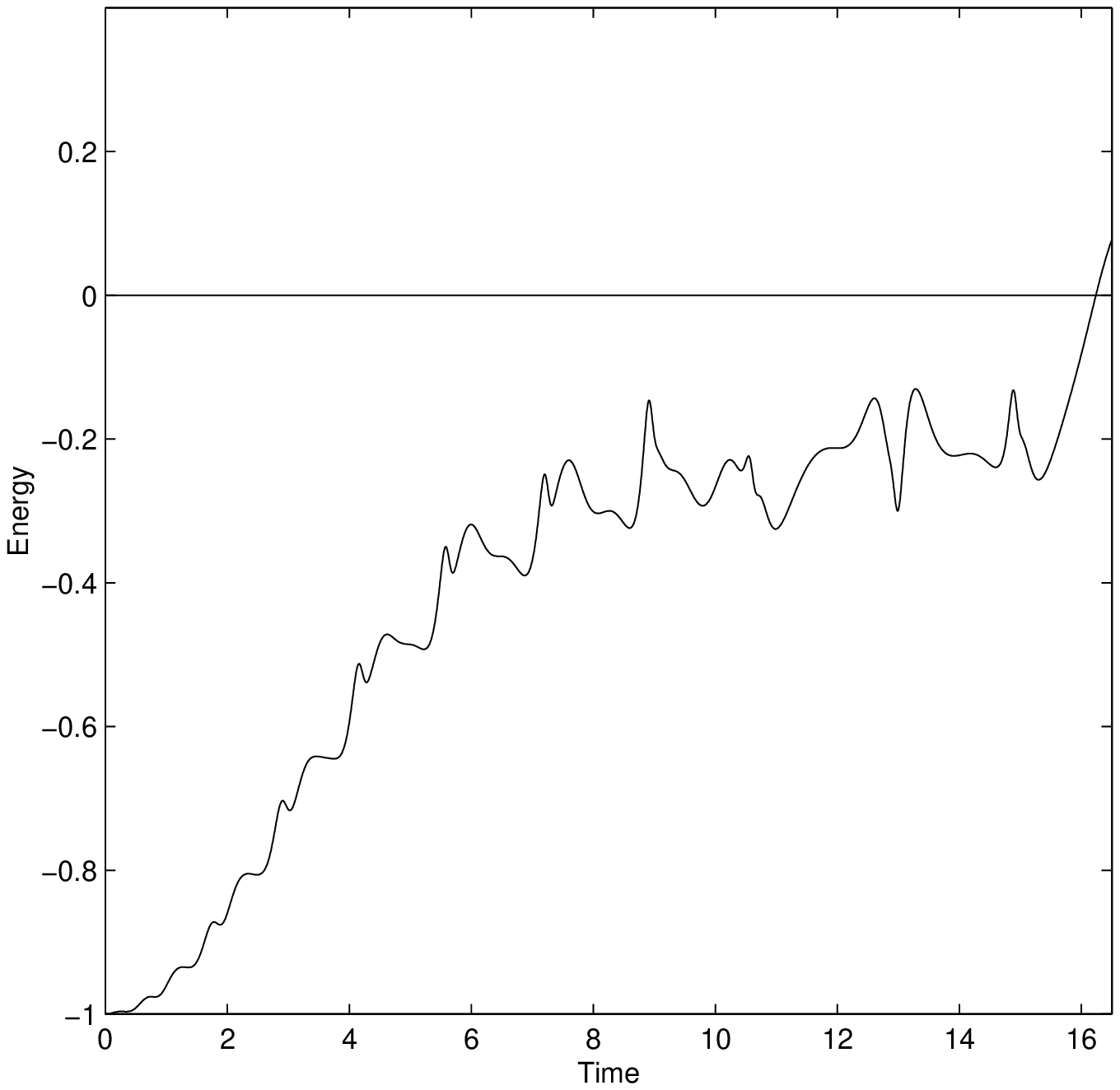}}
\caption{\label{fig-2} Excitation with linear frequency chirping:
 a) $\eps = 0.0157\,\frac{\w_0}{T_0}$,
 b) $\eps = 0.059\,\frac{\w_0}{T_0}$ (the chirping rate,
 which provides the fastest dissociation).
Energy and time are measured in units of $D$ and $T_0$.}
\end{figure}

\fig{2}a,b correspond to linear frequency chirping \eq{9} at two
different chirping rates. The dissociation is also achieved, but
it is slower then in the previous case and the results are very
sensitive to the variations of $\eps$. Additional flexibility to
the chirping process can added by using the relation
 $\w = \w_0 - \eps t$,
where $\w_0$ is a constant less then~$\W_0$. Optimal selection of
both parameters is a pretty complicated task. In more detail this
problem is investigated in~\cite{Liu95}.

\begin{figure}[htb]
a)\hspace{-1mm}
\hbox{\epsfbox{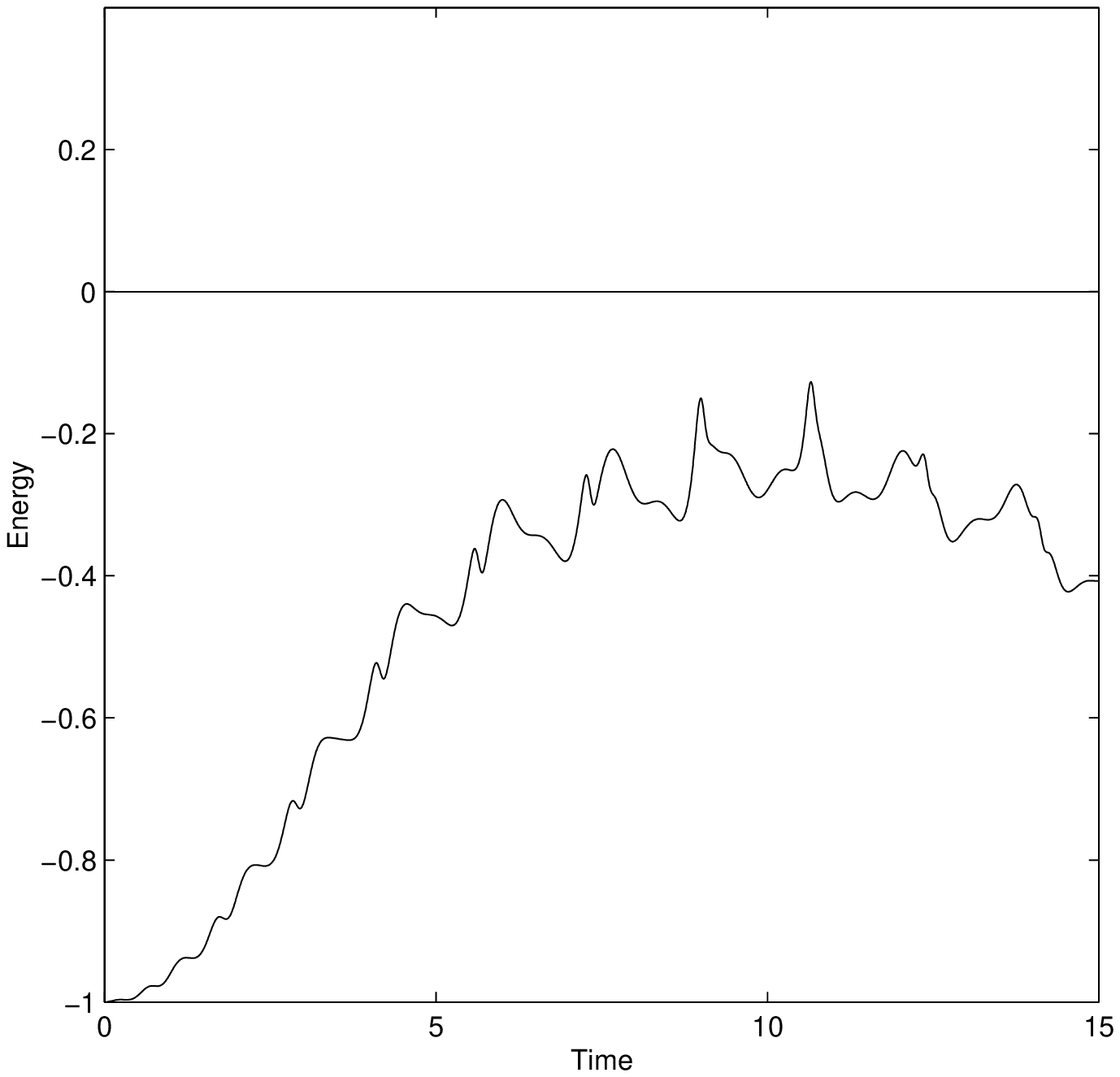}}
\hspace{7mm}
b)\hspace{-1mm}
\hbox{\epsfbox{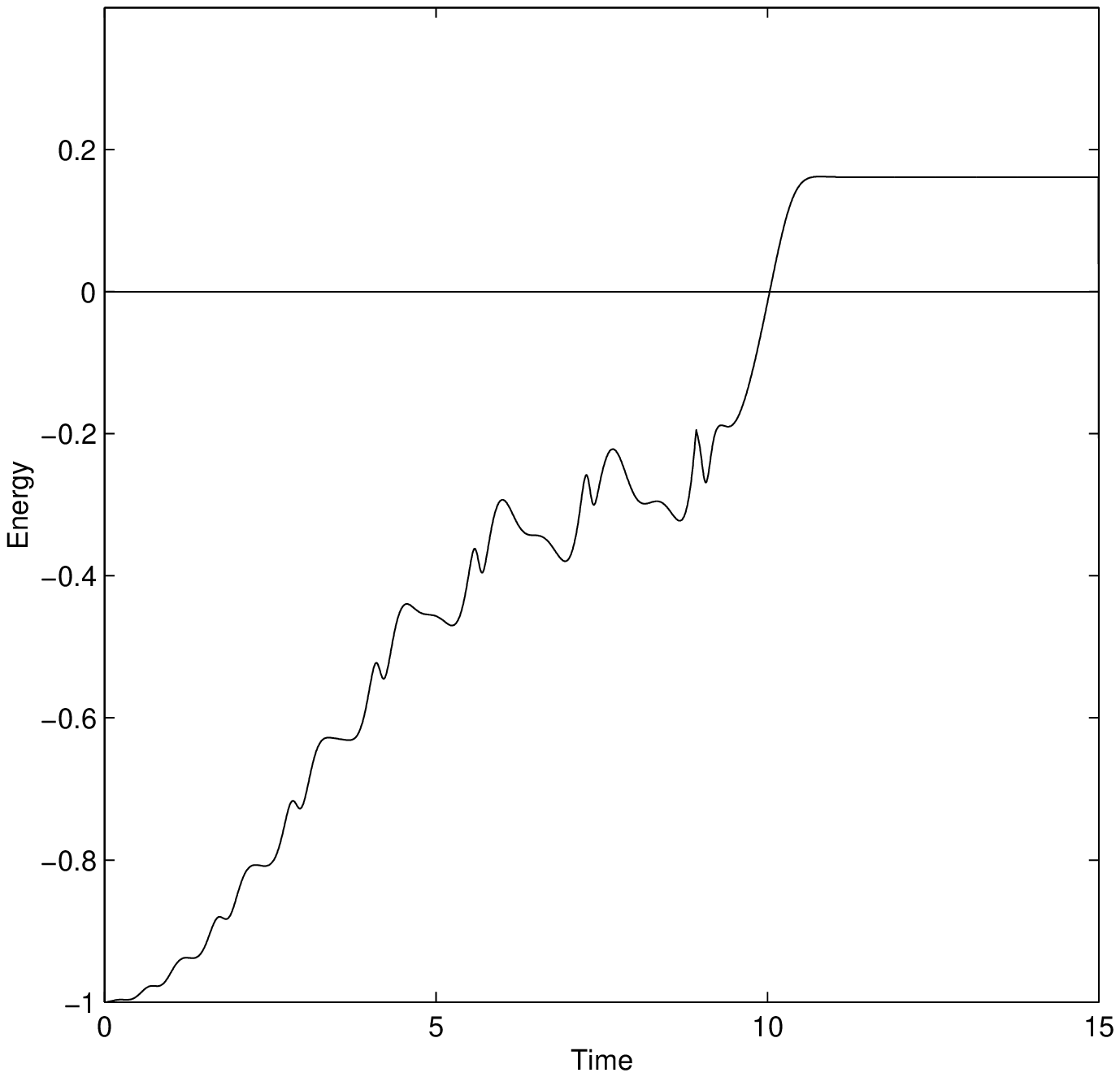}}
\caption{\label{fig-3}
Excitation with feedback control:
% a) $\dot\f = \W(W) = \W_0\sqrt{- W / D}$,
 a) using the frequency-energy relation; % $\dot\f = \W(W)$,
 b)~same, but with phase switch at $W = -0.2D$.
Energy and time are measured in $D$ and~$T_0$.}
\end{figure}

In this paper we will concentrate on the feedback chirping, based
on the frequency-energy curve \eq{14}. \fig{3}a shows the
energy as a function of time for the feedback control~\eq{15}. At
the initial stage the energy is growing faster then in other
cases, but after the level about $W=-0.5D$ is reached, the
efficiency of the excitation decreases so that dissociation is not
realized. Probably this phenomenon is caused by wrong phasing
which prevents the molecule dissociation although the resonance
condition is all the time fulfilled. Indeed, the sign of the
multiplier $\mu'(r)$ in equation \eq{3} due to formula \eq{6} is
changing for high energies, so that the excitation starts acting
in the opposite phase. To overcome this difficulty the following
algorithm can be used: when the energy of the molecule reaches
some certain value $W_*$ then the sign of the excitation force is
permanently changing to the opposite. The result of such an
excitation is shown in \fig{3}b. The dissociation is achieved at
$t = 10T_0$, which is faster then in all previous cases.

Thus for the selected level of control ($E = 0.1$ a.u.) the best
result is obtained for feedback excitation. However excitation
with shifted constant frequency gives the competitive result. The
situation changes if we consider lower level of excitation. If
this level is sufficiently small then excitation at any constant
frequency can not provide the dissociation. On the contrary,
excitation with changing frequency can in principle give
dissociation for any arbitrarily small level of control. For
illustration let us consider the 5 times smaller control then in
previous calculations: $E = 0.02$ a.u. In this case dissociation
at a constant frequency is not possible. Results of linear chirping
and feedback control excitation are given in \fig4. For simplicity
these calculations use approximation $\mu'(r)\ap\mu'(a)$. The
plots show that both methods provide dissociation. However, linear
chirping achieves dissociation at $t = 89 T_0$, while energy-feedback control
achieves it at $t = 34 T_0$.
Thus the feedback control provides more then two times
faster dissociation then the best variant of the linear chirping.

% but in the case
%of feedback control the dissociation is
%more then twice
%nearly tree times faster then for linear chirping:

\begin{figure}[htb]
a)\hspace{-1mm}
% \hbox{\epsfbox{fig06.eps}}
 \hbox{\epsfbox{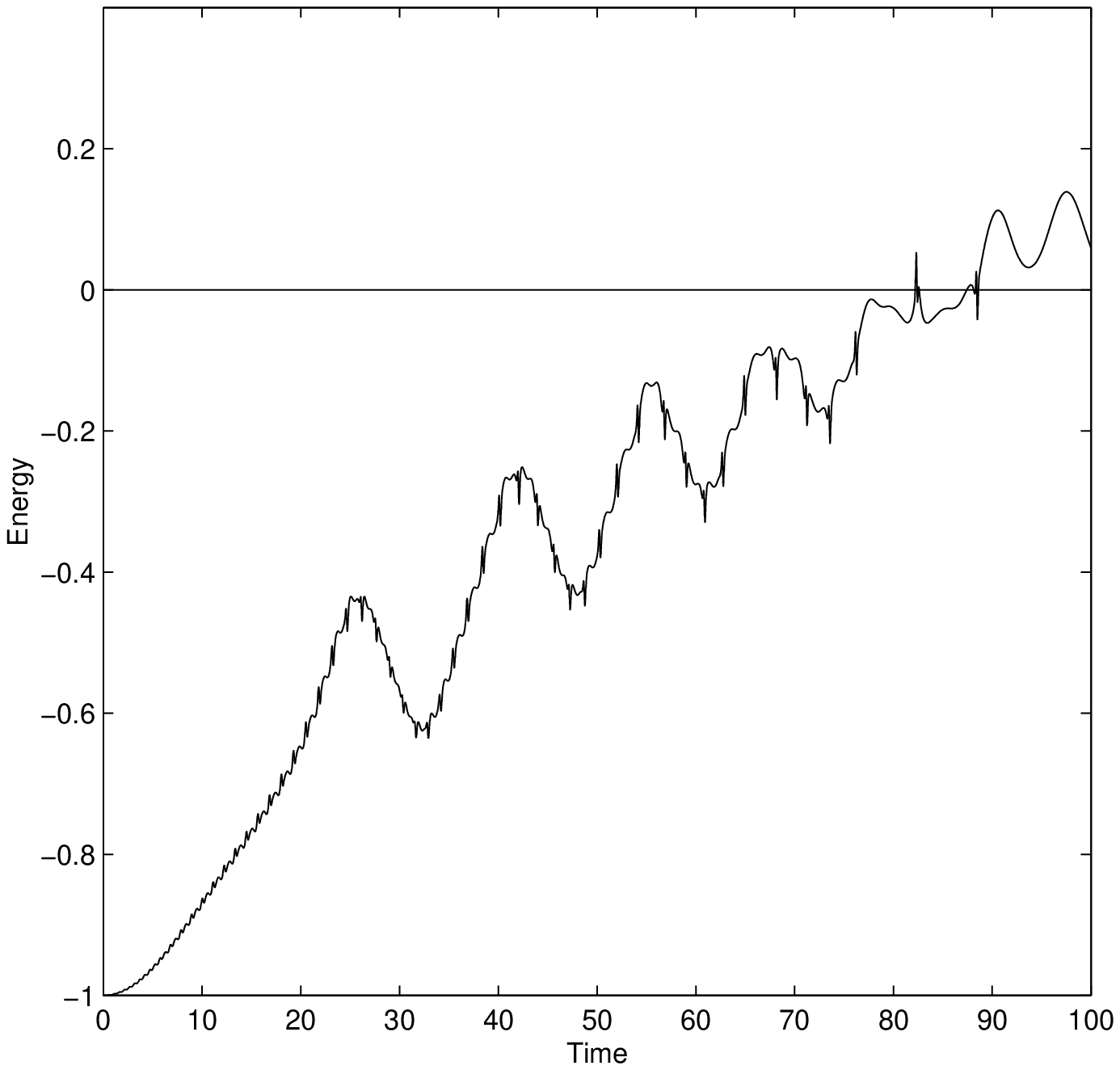}}
\hspace{7mm} b)\hspace{-1mm} \hbox{\epsfbox{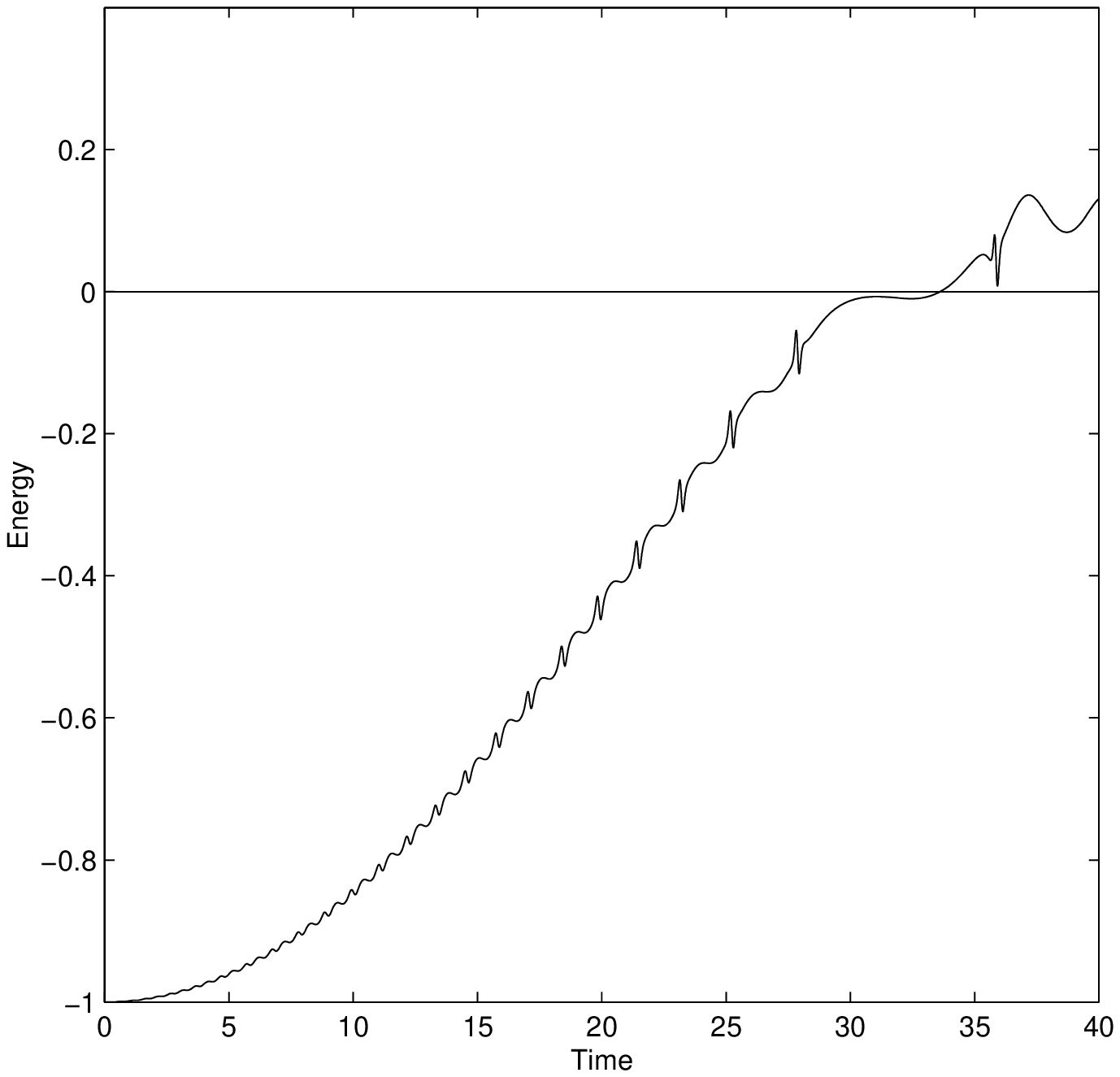}}
\caption{\label{fig-4} Excitation with lower level of control ($E
= 0.02$ a.u.) and constant value of dipole moment
% a) $\dot\f = \W(W) = \W_0\sqrt{- W / D}$,
% a) linear chirping with $\eps = 0.008\,\frac{\w_0}{T_0}$,
 a) linear chirping with $\eps = 0.0091\,\frac{\w_0}{T_0}$,
 b) feedback control (using frequency-energy relation). % $\dot\f = \W(W)$,
Energy and time are measured in $D$ and~$T_0$.}
\end{figure}

Although  the proposed algorithm requires measurement
of molecule energy only, its on-line implementation requires changing
 controlling action at the time intervals less then $10^{-12}$\,s.
 This can be achieved with modern ultrafast lasers. The approach
 can be easily generalized to control of an ensemble of molecules,
 replacing  exact energy of the separate molecule in the
 control objective by the average energy per single molecule.

%\section*{Acknowledgements}

\end{document}